\documentclass[12pt]{article}  
\usepackage{amssymb,latexsym}  
\usepackage{amsmath}  
 
\columnwidth\textwidth

\begin{document}

\newtheorem{df}{Definition}  
\newtheorem{thm}{Theorem}  
\newtheorem{lem}{Lemma}  
\newtheorem{rl}{Rule} 
\begin{titlepage}  
 
\noindent  
 
%\vspace*{3cm}  
\begin{center}  
{\LARGE Cluster separability, indispensability of detectors \\ and quantum
  measurement problem} 
\vspace{1cm}

P. H\'{a}j\'{\i}\v{c}ek \\  
Institute for Theoretical Physics \\  
University of Bern \\  
Sidlerstrasse 5, CH-3012 Bern, Switzerland \\  
hajicek@itp.unibe.ch

\vspace{1cm}  
 
%Preliminary version \\  
 
March 2010 \\  
\vspace{1cm}
 
 
\vspace*{2cm}  
 
\nopagebreak[4]  
 
\begin{abstract}
Careful analysis of cluster separability opens a way to a completely new
understanding of preparation and registration procedures for microsystems:
they are changes of separation status. An important observation is that
quantum mechanics does not specify what happens to systems when they change
their separation status. New rules that close this gap can therefore be added
without disturbing the logic of quantum mechanics. Another important
observation is that registration apparatuses for microsystems must contain
detectors and that their readings are signals from detectors. This leads to
further restrictions on measurability of observables, especially for
macroscopic quantum systems. Beltrametti-Cassinelli-Lahti model is used to
show how this approach leads to solution of the measurement problem. 
\end{abstract}

\end{center}

\end{titlepage}

Discussions about the nature of quantum measurement were started already by
founding fathers of the theory, persisted throughout and seem even to amplify
at the present time. An example is the quantum decoherence theory
\cite{zeh,zurek}, another the superselection sectors approach
\cite{hepp,primas}, etc. However, the problem is far from being satisfactorily
solved. For example, Refs. \cite{d'Espagnat,bub,BLM} analyse the well-known
shortcomings. We adopt the definition of the problem and the proof that it is
far from being solved from \cite{BLM} and give a short account of a new
approach to quantum measurement problem based on preliminary results of
\cite{PHJT,hajicek,hajicek2}. For the time being, it is applicable to
measurements on microscopic quantum systems. Large systems such as strong
laser signals or BECs will be studied later. 

The fact that particles of the same type are indistinguishable in quantum
mechanics implies that experiments with one particle are disturbed by another
particle of the same type, even if it were prepared independently, far away
from the first. 

Suppose that state $\psi(\vec{x}_1)$ of system ${\mathcal S}_1$ is prepared in
our laboratory as if no other system of this type existed. Next, let state
$\phi(\vec{x}_2)$ of system ${\mathcal S}_2$ of the same type be prepared
simultaneously in a remote laboratory. Then the state of the composite system
must be 
\begin{equation}\label{symst}
\Psi(\vec{x}_1,\vec{x}_2) =
\nu\bigl(\psi(\vec{x}_1)\phi(\vec{x}_2) \pm
\phi(\vec{x}_1)\psi(\vec{x}_2)\bigr)\ , 
\end{equation}
where $\nu$ is a normalisation factor. It follows that the state of ${\mathcal
  S}_1$ is not $\psi(\vec{x}_1)$ even if this state has been prepared in our
laboratory but that it depends on what has been done somewhere in the world in
another laboratory. 

Similarly, if the registration on ${\mathcal S}_1$ corresponding to an
observable $a(\vec{x}_1;\vec{x}'_1)$ of ${\mathcal S}_1$ is now performed in
our laboratory, it is equally possible that the registration is made on
${\mathcal S}_1$ or ${\mathcal S}_2$ and both can make a contribution to the
outcome. Hence, the correct observable that describes such a registration is
operator 
\begin{equation}\label{symop}
A(\vec{x}_1,\vec{x}_2;\vec{x}'_1,\vec{x}'_2) =
a(\vec{x}_1;\vec{x}'_1)\delta(\vec{x}_2-\vec{x}'_2) +
a(\vec{x}_2;\vec{x}'_2)\delta(\vec{x}_1-\vec{x}'_1)\ . 
\end{equation}

There seems to be no control of states that are prepared anywhere in the world
and the different possibilities have different measurable consequences. For
example, the position operator of ${\mathcal S}_1$ is $a(\vec{x}_1;\vec{x}'_1)
= \vec{x}_1\delta(\vec{x}_1-\vec{x}'_1)$ and suppose that the position is
registered. Then, the average in the prepared state is  
$$
\int d^3x_1\vec{x}_1\psi^*(\vec{x}_1)\psi(\vec{x}_1)\ .
$$
On the other hand, the preparation of ${\mathcal S}_2$ leads to a different
value: 
\begin{multline*}
\int d^3x_1d^3x_2 d^3x'_1d^3x'_2\Psi^*(\vec{x}'_1,\vec{x}'_2)
A(\vec{x}_1,\vec{x}_2;\vec{x}'_1,\vec{x}'_2) \Psi(\vec{x}_1,\vec{x}_2) \\ =
\int d^3x_1\vec{x}_1\psi^*(\vec{x}_1)\psi(\vec{x}_1) + \int d^3x_1\vec{x}_1
\phi^*(\vec{x}_1)\phi(\vec{x}_1)\ .  
\end{multline*}

Cluster separability (see, e.g., \cite{peres}, P. 128) was invented to deal
with these problems. The key notion is that of $D$-local observable: 
\begin{df}
Let $a(\vec{x}_1;\vec{x}'_1)$ be an observable of ${\mathcal S}_1$, let $D$ be
the domain of $\vec{x}_1$ inside our laboratory and let  
\begin{equation}\label{local1}
\int d^3x_1a(\vec{x}_1;\vec{x}'_1)f(\vec{x}_1) = \int
d^3x'_1a(\vec{x}_1;\vec{x}'_1)f(\vec{x'}_1) = 0  
\end{equation}
if (supp\,$f)\cap D = \emptyset$, where $f$ is a test function. Let us call
such operators $D$-{\em local}.  
\end{df}

Let us assume that (supp\,$\psi) \subset D$ and (supp\,$\phi) \cap D =
\emptyset$. If ${\mathcal S}_2$ has been prepared and the $D$-local observable
$a_D(\vec{x}_1,\vec{x}'_1)$ is used instead of $a(\vec{x}_1;\vec{x}'_1)$ in
formula (\ref{symop}) defining operator ${\mathsf A}_D$ instead of ${\mathsf
  A}$ and we obtain  
\begin{multline*}
\int_D d^3x_1\int_Dd^3x'_1\int_Dd^3x_2\int_Dd^3x'_2\Psi^*(\vec{x}_1,\vec{x}_2)
A_D(\vec{x}_1,\vec{x}'_1;\vec{x}_2,\vec{x}'_2)\Psi(\vec{x}'_1,\vec{x}'_2) \\ =
\int_{-\infty}^\infty d^3x_1\int_{-\infty}^\infty
d^3x'_1\psi^*(\vec{x}_1)a(\vec{x}_1;\vec{x}'_1)\psi(\vec{x}'_1)
\end{multline*} 
as if no ${\mathcal S}_2$ existed. It follows that in this case both rules for
individual systems and rules for identical subsystems lead to the same
results. 

However, observables that are usually associated with ${\mathcal S}_1$ are not
$D$-local. For example, the position operator
$\vec{x}_1\delta(\vec{x}_1-\vec{x}'_1)$ violates the condition by large
margin, as seen above. In fact, this 'observable' controls position of the
system in the whole infinite space. This is utterly different from observables
that can be registered in a human laboratory. Nevertheless, one can construct
observables for ${\mathcal S}_1$ that are $D$-local and are, therefore, more
realistic (see \cite{hajicek2}). 

This motivates introduction of the concept of {\em separation status} of a
microsystem. For example, a microsystem that is alone in the Universe has a
different separation status than the same microsystem in a domain $D$ in which
there are no microsystem of the same type but which is surrounded by matter
containing a lot of such microsystems (assuming, of course, that supports of
their states do not intersect $D$). Observables of the first system are the
standard ones, whereas observables of the second are only the $D$-local
ones. An extreme case of separation status for a microsystem is if its 'free'
domain $D$ shrinks to zero and the only available modus of description for it
is that of identical subsystems. 

We observe that the current version of quantum mechanics is a theory of
systems with a fixed separation status. Let us call this restricted
understanding {\em fixed-status quantum mechanics} (FSQM). It deals with
single microsystems as if no other microsystems of the same type existed
according to one set of rules and with composite systems containing many
identical subsystems according to another set of rules. It neglects the
obvious relations that make such an approach in principle inconsistent. Still,
the method works and the justification why it works is the cluster
separability. However, FSQM might have limits and the limits have to do with
changes of the separation status of a system during its time evolution. 

Next, we have to look at the registration apparatuses. A careful study leads
to the following observation: 
\begin{rl}
Any registration apparatus for microsystems must contain at least one detector
and every reading of an apparatus value is a signal from a detector. 
\end{rl}
Rule 1 seems to be obvious but it has far reaching consequences for existence
of observables. 

An important example are observables of macroscopic systems. In general, a
macroscopic system ${\mathcal A}$ is a composite quantum system with very many
different microsystem constituents. One can subdivide these microsystems into
type classes. Consider first observables that concern properties of
microscopic subsystem ${\mathcal S}$ of type $\tau$. If we apply the basic
rules of observable construction for systems of identical microsystems, then
e.g.\ the position and momentum of any individual microsystem ${\mathcal S}$
are not observables of ${\mathcal A}$ but 'collective' one-particle operators
such as 
$$
{\mathsf a}_{\text{col}} = \sum_k a(\vec{x}_k;\vec{x}'_k)\ ,
$$
 could be, where $a(\vec{x}_k;\vec{x}'_k)$ is an operator acting on $k$-th
 subsystem of type $\tau$. Suppose that there is an apparatus ${\mathcal B}$
 suitable to measure $a(\vec{x};\vec{x}')$ on an individually prepared system
 ${\mathcal S}$. One can imagine that applying ${\mathcal B}$ to ${\mathcal
   A}$ in some way similar to that described above would measure ${\mathsf
   a}_{\text{col}}$ because any subsystem of type $\tau$ would automatically
 contribute to the result of the measurement. However, the registration
 apparatus was applied to individually prepared subsystems there. It follows
 from Rule 1 that the apparatus ${\mathcal B}$ cannot be applied to ${\mathcal
   A}$ in this way because none of the identical subsystems of type $\tau$ are
 prepared individually. Its readings are signals of its detector and for any
 detector to register ${\mathcal S}$, ${\mathcal S}$ must be isolated to be
 manipulable, have sufficient kinetic energy, etc. Hence, to measure
 collective observable ${\mathsf a}_{\text{col}}$, we need a method that makes
 measurements directly on ${\mathcal A}$. 

For example, let ${\mathcal A}$ be a crystal. By scattering $X$-rays off it,
relative positions of its nuclei can be recognised. But rather than a position
of individual nucleus it is a space dependence of the average nuclear density
due to all nuclei. Such an average nuclear density could be obtained with the
help of an operator similar to ${\mathsf a}_{\text{col}}$. In general,
scattering of a microsystem ${\mathcal S}'$ off a macrosystem ${\mathcal A}$
can be calculated from potential $V_k(\vec{x},\xi, \vec{x}_k,\xi_k)$ that
describes the interaction between ${\mathcal S}'$ and one of the microscopic
subsystems of ${\mathcal A}$. The whole interaction Hamiltonian is then a sum
extending over all subsystems that interact with ${\mathcal S}'$, 
\begin{equation}\label{coll}
\sum_k V_k(\vec{x},\xi, \vec{x}_k,\xi_k)\ .
\end{equation}
It is important to realise that there are very few interactions and not all of
them can supply potentials useful for practical experiment. 

Another example is the kinetic energy of ${\mathcal S}$. Again, the
corresponding collective observable cannot be measured by the method kinetic
energy is measured on individual systems of type ${\mathcal S}$. But the
average of the collective observable could have the meaning of $N/k_B$ times
the temperature, where $N$ is the number of ${\mathcal S}$-constituents of
${\mathcal A}$. Hence, a viable method to measure the average is to measure
the temperature of ${\mathcal A}$. Again, this is a very special case that
works only under specific conditions. Further examples have to do with other
additive quantities, such as momentum and angular momentum. Total values of
these quantities can be measured and they are of the form (\ref{coll}). 

We notice, first, that an observable $a(\vec{x}_k;\vec{x}'_k)$ of a
microsystem ${\mathcal S}$ can be promoted to a collective observable
${\mathsf a}_{\text{col}}$ of ${\mathcal A}$ if ${\mathcal A}$ admits a direct
measurement of ${\mathsf a}_{\text{col}}$, which happens only in special and
rare cases. Second, such a collective observable is still too 'sharp', because
only some averages with rather large variances can be observed. It is
impossible to obtain its single eigenvalues as results of registration (for an
example, see Ref.\ \cite{peres}, P.181). Thus, one of the consequences of Rule
1 is a principal and severe limit on mathematically well-defined quantities
being observables. For more discussion, see \cite{hajicek2}. 

The idea that the fixed-status quantum mechanics (FSQM) has obvious limits as
well as the idea of detector indispensability lead to a considerable
modification of quantum theory of measurement. The necessary changes are:  
\begin{enumerate}
\item Preparation has a different and much greater significance than is
  usually assumed. Any preparation gives the prepared system ${\mathcal S}$
  its objective quantum properties such as states, gemenge structures,
  averages and variances of observables (for extended discussion, see
  \cite{PHJT}) so that it is justified to speak of a physical object. This is
  what we call {\em quantum object}. Simultaneously, a preparation must
  separate ${\mathcal S}$ from the set of identical microsystems, at least
  approximately. The prepared state must be $D$-local in a suitable domain
  $D$. Only then, ${\mathcal S}$ can be viewed as an individual system and the
  standard notion of observable becomes applicable to it. This is justified by
  the idea of cluster-separability. Finally, a preparation may isolate the
  microsystem so that it can be individually manipulated by e.g.\ external
  fields or mater shields and registered by detectors. 
\item Registration has a more specific form than is usually assumed. Any
  apparatus that is to register a microsystems directly contains a detector
  and the 'pointer' value that is read off the apparatus is a signal from the
  detector. We assume that each detector contains a bulk of sensitive matter
  (see, e.g., \cite{leo}) with which the registered microsystem is unified and
  changes its separation status. The change of separation status during a
  registration is a similar to, but a deeper change than, the so-called
  collapse of the wave function. Indeed, standard quantum mechanics provides
  no information about processes such as preparation and registration, in
  which the separation status of microsystems changes. Unjustified
  applications of standard rules to such processes leads to contradictions
  with experimental evidence (the so-called objectification problem, see
  \cite{BLM}). However, one can add new rules to quantum mechanics governing
  such processes without violating its logic. Macrosystem ${\mathcal A}$ such
  as a blocking shield, a scattering target or a detector that contains
  microsystems indistinguishable from ${\mathcal S}$ must lie at the boundary
  of $D$. Corrections to FSQM description of the behaviour of the composed
  system ${\mathcal S} + {\mathcal A}$ due to a possible separation status
  change of ${\mathcal S}$ must be carefully chosen.  
\end{enumerate}
The usual method of FSQM is to specify initial states of both ${\mathcal S}$
and ${\mathcal A}$ before their interaction, choose some appropriate
interaction Hamiltonian and calculate the corresponding unitary evolution of
the composed system ${\mathcal S} + {\mathcal A}$ ignoring the problem with
separation status change. We shall now show an example of corrections that
must be done. 

Let a discrete observable ${\mathsf O}$ of system ${\mathcal S}$ be
measured. Let $o_k$ be eigenvalues and $\{\phi_{kj}\}$ be the complete
orthonormal set of eigenvectors, ${\mathsf O}\phi_{kj} = o_k \phi_{kj}$. Let
the registration apparatus be a quantum system ${\mathcal A}$. Let ${\mathsf
  A}$ be a non-degenerate, discrete observable of ${\mathcal A}$ with the same
eigenvalues $o_k$ and with the complete orthonormal set of eigenvectors
$\psi_k$, ${\mathsf A}\psi_k = o_k \psi_k$. 

Let the measurement start with the preparation of ${\mathcal S}$ in state
$\phi$ and the independent preparation of ${\mathcal A}$ in state $\psi$. Let
${\mathcal S}$ and ${\mathcal A}$ then interact for a finite time and let the
resulting state be given by ${\mathsf U}(\phi \otimes \psi)$, where ${\mathsf
  U}$ is a unitary transformation. Unitary evolution of $\phi \otimes \psi$
and discreteness of ${\mathsf O}$ are definition properties of
Beltrametti-Cassinelli-Lahti model of premeasurement \cite{belt}. Then, for
any initial state $\psi$ of ${\mathcal A}$, there is a set $\{\varphi_{kl}\}$
of unit vectors in the Hilbert space of ${\mathcal S}$ satisfying the
orthogonality conditions $\langle \varphi_{kl}|\varphi_{kj}\rangle =
\delta_{lj}$ such that ${\mathsf U}$ is a unitary extension of the map 
\begin{equation}\label{unitar}
\phi_{kl}\otimes \psi \mapsto \varphi_{kl}\otimes \psi_k\ .
\end{equation}
For proof, see \cite{belt}.

To obtain a model of measurement, it is necessary (but not sufficient) that 
\begin{equation}\label{measur}
\langle \varphi_{ki}|\varphi_{lj}\rangle = \delta_{kl}\delta_{ij}\ .
\end{equation}
Let the initial state of ${\mathcal S}$ be an arbitrary state $\phi$.
Decomposing $\phi$ into the eigenstates,
$$
\phi = \sum_{kl} c_{kl}\phi_{kl}\ ,
$$
we obtain from Eq.\ (\ref{unitar}) 
\begin{equation}\label{finalSA}
{\mathsf U} (\phi \otimes \psi) = \sum_k \sqrt{p^{\mathsf
    O}_\phi(o_k)}\Phi_k\otimes \psi_k\ ,  
\end{equation}
where
$$
\Phi_k = \frac{\sum_l c_{kl}\varphi_{kl}}{\sqrt{\langle \sum_l
    c_{kl}\varphi_{kl}|\sum_j c_{kj}\varphi_{kj}\rangle}}  
$$
and 
$$
p^{\mathsf O}_\phi(o_k) = \left\langle \sum_l c_{kl}\varphi_{kl}\Biggm|\sum_j
  c_{kj}\varphi_{kj}\right\rangle  
$$
is the probability that a registration of ${\mathsf O}$ performed on vector
state $\phi$ gives the value $o_k$. The final state of apparatus ${\mathcal
  A}$ then follows from Eq.\ (\ref{measur}), 
\begin{equation}\label{gemengA}
tr_{\mathcal S}[|{\mathsf U}(\phi \otimes \psi)\rangle\langle {\mathsf U}(\phi
\otimes \psi)|] = \sum_j p^{\mathsf O}_\phi(o_j)|\psi_j\rangle\langle\psi_j|\
. 
\end{equation}

At the end of the registration, ${\mathcal A}$ must objectively be in one of
the states $\psi_j$ in each individual case (objectification requirement).
That is, the right-hand side of Eq.\ (\ref{gemengA}) must be the {\em gemenge
  structure} of the state \cite{BLM,hajicek2} (some authors \cite{d'Espagnat}
use the term 'proper mixture' instead of 'gemenge'). However, state
(\ref{gemengA}) of ${\mathcal A}$ has not the gemenge structure given by the
right-hand side of Eq.\ (\ref{gemengA}) because of the entanglement with
${\mathcal S}$ due to state (\ref{finalSA}). The reason is that state
(\ref{finalSA}) contains much more correlations between observables of
${\mathcal S}$ and ${\mathcal A}$ than just correlations between the states
$\Phi_k$ and $\psi_k$. To measure any of these correlations, we would always
need some observables of ${\mathcal S}$ that do not commute with ${\mathsf
  O}$.

However, the assumption that the end state of ${\mathcal S} + {\mathcal A}$ is
(\ref{finalSA}) seems to be an illusion. Microsystem ${\mathcal S}$ is assumed
to be somewhere inside ${\mathcal A}$ at this stage and is indistinguishable
from other microsystems of the same type within ${\mathcal A}$. There is
always a lot of them, either because they are present in the detector(s)
before the registration started or because the detector(s) becomes quickly
polluted by them afterwards. Thus, the separation status of the system
${\mathcal S}$ has changed and with it also the separation status of the whole
composite system ${\mathcal S} + {\mathcal A}$ has. The applications of FSQM
to two systems of different separation status is different. In our case,
system ${\mathcal S} + {\mathcal A}$ before the interaction is a composite one
and each of the subsystems is an object having its states and observables.
During and after the interaction, however, ${\mathcal S}$ ceases to be an
object, becomes a part of ${\mathcal A}$ and looses all of its observables
except of ${\mathsf O}$. This is a deeper change than just a change of state.
Hence, the existence of most correlations that define state (\ref{finalSA}) is
lost. The point is not that some observables are difficult to measure but
rather that these observables do not exist at all. The only correlations that
can remain are those between end states $\psi_k$ of ${\mathcal A}$ and
$\Phi_k$ of the microsystem. They define the state
\begin{equation}\label{truend}
\sum_k p^{\mathsf O}_\phi(o_k)|\Phi_k\rangle \langle\Phi_k| \otimes
|\psi_k\rangle 
\langle\psi_k|\ . 
\end{equation}
This motivates the following assumption: 
\begin{rl}
Let discrete observable ${\mathsf O}$ of microsystem ${\mathcal S}$ be
registered by apparatus ${\mathcal A}$ and the 
corresponding unitary evolution leads to the state (\ref{finalSA}) with
${\mathcal S}$ inside ${\mathcal A}$. Then the true state of ${\mathcal S} +
  {\mathcal A}$ and its gemenge structure are given by Eq.\
  (\ref{truend}).
\end{rl}
 The end state of ${\mathcal A}$ has then necessarily gemenge structure
 (\ref{gemengA}). The content of Rule 2 is that only the correlations between
 the states $\psi_k$ of ${\mathcal A}$ and $\Phi_k$ of ${\mathcal S}$ 
survive and all other correlations between ${\mathcal A}$ and
${\mathcal S}$ are erased during the change of separation status of ${\mathcal
  S} + {\mathcal A}$. What survives and what is erased is uniquely determined
by the Beltrametti-Cassinelli-Lahti model. In particular, states
$\varphi_{kl}$ are uniquely determined by initial state $\psi$ of 
${\mathcal A}$ and initial state $\phi$ of
${\mathcal S}$ determines states $\Phi_k$ uniquely. Thus, the additional
evolution from state (\ref{finalSA}) to state (\ref{truend}) is non-unitary
but still deterministic. Rule 2 is a new general rule which has to be
added to quantum mechanics. To choose such a rule, we have to look at
observations and experiments. Rule 2 is in an agreement with what is observed.
For more discussion of Rule 2, see \cite{hajicek2}.

It ought to be clear from this example how our method works. Further possible
questions are discussed in \cite{hajicek2}.

\end{document}